\documentstyle[twoside,fleqn,espcrc2,epsf]{article}

\title{  Re ${ A_0}$ and Re ${ A_2}$ from Quenched Lattice QCD}
\epsfclipon
 
\author{Calin R. Cristian\address{Department of Physics, Columbia
  University, New York, NY, 10027, USA}\thanks{This work was done in 
collaboration with T. Blum, N. Christ,
  C. Dawson, G. Fleming, X. Liao, G. Liu, R. Mawhinney, S. Ohta, A. Soni,
  P. Varanas, M. Wingate, L. Wu, Y. Zhestkov. We thank RIKEN,
  Brookhaven National Laboratory and the U.S. Department of Energy for
  providing the facilities essential for this work.}\\ 
  RBC Collaboration
}
 
\begin{document}

\begin{abstract}

We have used domain wall fermions to calculate $K \rightarrow \pi$ and 
$K \rightarrow 0$ matrix elements which can be used to study the 
$\Delta I = 1/2$ rule for K decays in the Standard Model. Nonlinearities in
the $\Delta I = 3/2$ matrix elements due to chiral logarithms are 
explored and the subtractions needed for the $\Delta I = 1/2$ matrix 
elements are discussed. Using renormalization factors calculated using
non-perturbative renormalization then yields values for real $A_0$ and
$A_2$. We present the details of our quenched $16^3 \times 32 \times
16$, $\beta = 6.0$, $M_5 = 1.8$ simulation, where a previous
calculation showed that the finite $L_s$ chiral symmetry breaking 
effects are small ($m_{\rm res} \approx 4 \rm MeV$). 

\end{abstract}

\maketitle
\section{INTRODUCTION}
At energies below the electroweak scale the weak interactions are 
described by local four-fermi operators multiplied by effective
coupling constants, the Wilson coefficients. The formal framework to
achieve this is the operator product expansion (OPE) which allows one to
separate the calculation of a physical amplitude into two distinct
parts: the short distance (perturbative) calculation of the Wilson
coefficients and the long distance (generally non-perturbative)
calculation of the hadronic matrix elements of the operators $Q_i$. 
We calculate on the lattice $K \rightarrow \pi$ and $K \rightarrow 0$. 
This allows us to calculate the low energy constants in chiral
perturbation \cite{bernard} which, after incorporating the
non-perturbative renormalization factors are then translated into 
$K \rightarrow \pi \pi$ matrix elements. The CP-PACS collaboration has
also presented a very similar calculation at this meeting \cite{noaki}. 

\section{DETAILS OF THE SIMULATIONS}
We have used the Wilson gauge action, quenched, at $\beta = 6.0$  on a
$16^3 \times 32$ lattice which corresponds to an inverse lattice 
spacing $a^{-1}=1.922 \, {\rm GeV}$. 
The domain wall fermion height $M_5 = 1.8$ and fifth dimension $L_s
=16$ give a residual symmetry breaking  $m_{\rm res} = 0.00124 \; \approx
4 \, {\rm  MeV}$ \cite{dwf_hadron_00}; 400 
configurations separated by 10000 heat-bath sweeps were used in this analysis. 
$K \rightarrow \pi$ matrix elements were calculated 
in the $SU(3)_{\rm flavor}$ limit for 5 light quark masses $m_f = 0.01,
0.02, 0.03, 0.04, 0.05$. Since the $K \rightarrow 0$ matrix elements 
vanish in the $SU(3)$ limit these matrix elements were calculated with
non-degenerate quark propagators for 10 mass combinations subject to
the constraint $m_s - m_d = 0.01, 0.02, 0.03, 0.04$. 
We have also calculated the so called eye diagrams with an active
charm quark for $m_f = 0.1, 0.2, 0.3, 0.4$ ( the physical charm quark
is around 0.5). However, the analysis for charm-in is still in
progress; in this presentation we concentrate on the case with 
3-active flavors wherein charm is integrated out assuming it is very heavy.
The calculation took about 4 months on 800 Gflops (peak). Quark
propagators were calculated using the conjugate gradient method with a stopping
residual of $10^{-8}$ with periodic and anti-periodic boundary
conditions which amounts to doubling the lattice size in time
direction. The two wall source propagators at $t_K=5$
and $t_{\pi}=27$  were fixed to Coulomb gauge.
For eye diagrams we employed random wall sources spread over time 
slices $t = 14 - 17$ with 2 hits per configuration.
Dividing the three-point correlation functions by the wall-wall 
pseudoscalar-pseudoscalar correlation function yields the
desired matrix elements up to a factor of $2 m_\pi$ which is determined from a
covariant fit to the wall-point two-point function in the range 
$t = 12 - 20$ for each mass.
%We have performed numerous checks, the most important being an 
%independent code written to check the contractions. 
%We have also repeated every tenth configuration. 
\section{CALCULATION OF LOW ENERGY CONSTANTS}
\begin{figure}[htb]
\epsfxsize = \hsize
\epsfbox{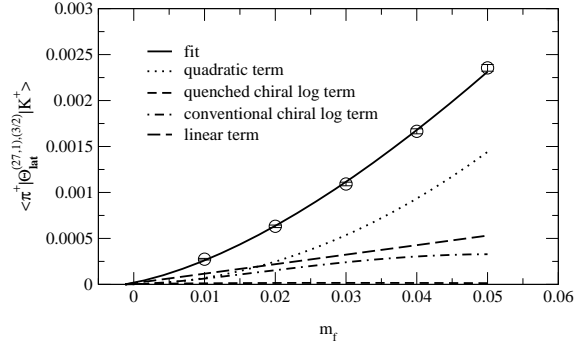}
\vskip -0.3in
\caption{The matrix element 
$\langle \pi^+ |\Theta_{lat}^{(27,1)} | K^+ \rangle $.}
\label{fig:o2_ktopi_3_2}
\vskip -0.2in
\end{figure}

\begin{figure}[htb]
\vskip -0.1in
\epsfxsize=\hsize
\epsfbox{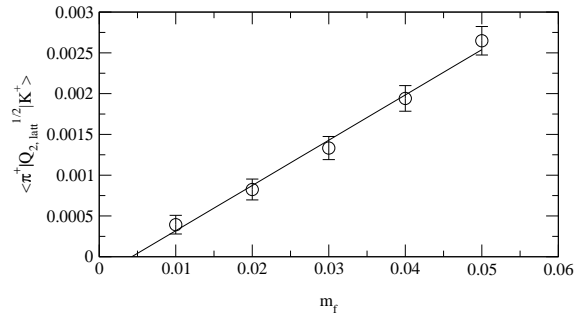}
\vskip -0.3in
\caption{The matrix element $\langle \pi^+
| Q_{i}^{1/2} | K^+ \rangle_{sub}$ }
\label{fig:o2_ktopi_1_2}
\vskip -0.2in
\end{figure}

Since our results unambiguously show that Re$A_0$ and Re$A_2$ come
essentially from the current-current operators (recall these have the 
largest Wilson coefficients) we will concentrate on
these operators from now on.
Quenched chiral perturbation theory predicts 
%\begin{small}
%\begin{equation}
$$
m_{\pi}^2 = 
a_\pi ( m_f + m_{\rm res} ) \left[ 1 - \delta 
\ln \left( \frac{ a_\pi ( m_f + m_{\rm res} )}
{\Lambda_{Q \chi pt}^2} \right) \right]    
%\end{equation}
%\end{small}
$$
We find a quenched chiral logarithm coefficient $\delta = 0.05(2)$ which has a 
negligible contribution in our matrix element calculation. Unlike the quenched
chiral logarithms, the conventional logarithms coming from quenched
chiral perturbation theory induce large corrections to the 
$\Delta I = 3/2$ $K \rightarrow \pi$ matrix element as can be
seen in Figure \ref{fig:o2_ktopi_3_2}. We fit these amplitudes to 
\cite{golterman} 
%\begin{equation}
$$
b_1^{(27,1)} m_M^2 [ 1 - ( \delta + \frac{6 m_M^2}{(4 \pi f)^2})
 \ln (\frac{m_M^2}{\Lambda^2} ) ] + b_2^{(27,1)} m_M^4
%\end{equation}
$$
where $m_M^2 = a_\pi (m_f+m_{res})$, $f=137 \, {\rm MeV}$, 
$\Lambda = 1 \, {\rm GeV}$. The conventional chiral logarithm 
$m_M^2 \ln (m_M^2 / \Lambda^2)$ is almost linear over the mass range
we have used so the fitting routine cannot distinguish this term 
from the linear term if we leave the coefficient of the logarithm as a free
parameter. Since the large coefficient -6 of the logarithm makes the
contribution of this term comparable to the contribution of the
linear term omitting this term would change
$b_1^{(27,1)}$ by almost a factor of two. The quenched chiral log
contribution is very small.
$\Delta I = 1/2$ $K \rightarrow \pi$ matrix elements mix with
$\bar{s}d$ with a power divergent coefficient $\sim (m_s + m_d)/a^2$.
We define a subtracted matrix element $\langle \pi^+ | Q_{i}^{(8,1)} |
K^+ \rangle_{sub}$ by
%\begin{equation}
$$
\langle \pi^+ | Q_{i}^{(8,1)} | K^+ \rangle + 
\eta_i (m_s + m_d) \langle \pi ^+ | \bar{s}d| K^+ \rangle
$$
%\end{equation}
where $\eta_i$ is obtained from a linear fit to 
$\frac{\langle 0 | Q_i | K \rangle}{\langle 0 | \bar{s} \gamma^5 d | K
\rangle }$. 
For an explanation of this subtraction of the power divergence 
we refer the reader to \cite{dwf_wme_00}. 
The quenched chiral perturbation theory corrections to  $\langle \pi^+
| Q_{i}^{1/2} | K^+ \rangle_{sub}$ are not known. As can be seen in
Figure \ref{fig:o2_ktopi_1_2} our data is consistent with a 
linear fit $c_{0,i} + c_{1,i}m_f$ with the slope $c_{1,i}$ determining 
the low energy constants $\alpha_i^{1/2}$ and the intercept $c_{0,i}$ 
arising from residual chiral symmetry breaking.
\section{ Re ${ A_0}$ AND Re ${ A_2}$ }
\begin{figure}[htb]
\vskip -0.1in
\epsfxsize=\hsize
\epsfbox{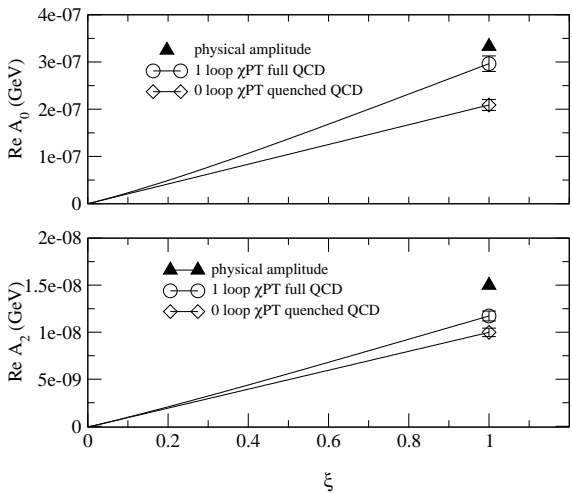}
\vskip -0.3in
\caption{ Re${ A_0}$ and  Re${ A_2}$ as a function of $\xi$}
\label{fig:re_a0_re_a2}
\vskip -0.2in     
\end{figure}

\begin{figure}[htb]
\epsfxsize=\hsize
\epsfbox{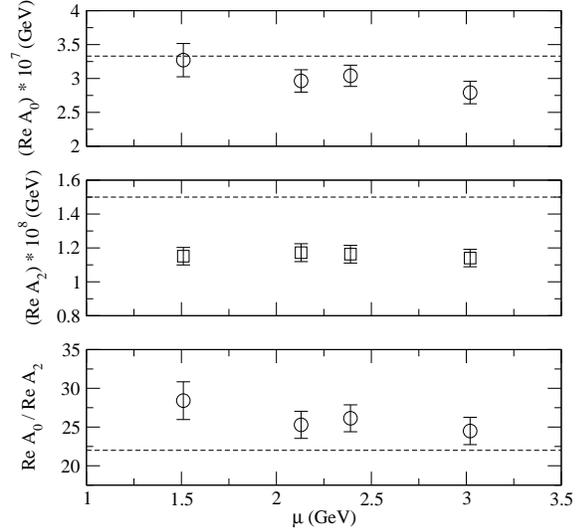}
\vskip -0.3in
\caption{ Re${ A_0}$ and  Re${ A_2}$ obtained using 1-loop logarithmic
corrections vs. the matching scale $\mu$; the dotted lines show the 
experimental results. }
\label{fig:delta_i_1_2_mu_dep}
\vskip -0.2in
\end{figure}

We use chiral perturbation theory to compute the lattice $K \rightarrow \pi
\pi$ matrix elements. Using non-perturbative Z factors we obtain the 
continuum matrix elements which are then multiplied by Wilson coefficients
to yield the physical amplitudes. We present an extrapolation to the
Kaon mass scale to lowest order in chiral 
perturbation theory and a second
extrapolation which includes one loop logarithmic effects.
We multiply the pseudoscalar masses by $\xi$ so that 
for $\xi \rightarrow 0$ the chiral perturbation theory extrapolation 
is increasingly accurate but we need the extrapolation at $\xi=1$, 
the physical point. 
In figure \ref{fig:re_a0_re_a2} we present Re${ A_0}$ and Re${ A_2}$
as a function of the parameter $\xi$.
The chiral logarithm correction for Re${ A_0}$ is large
(about $40\%$). In addition one expects a large correction (not
included here) coming from the
tree level $O(p^4)$ terms necessary to cancel the dependence on the
chiral perturbation theory scale $\Lambda_{\chi PT}$. 
If the Z factors and the Wilson coefficients were calculated to all
orders in perturbation theory the physical amplitudes that we
calculate would not depend on the scale $\mu$ where the
transition between the lattice and the continuum operators is made. To
a good approximation this is what we find, even though 
at $\mu = 1.51 \, {\rm GeV}$ one expects non-perturbative effects 
in the Z factors and at $\mu=3.02 \, {\rm GeV}$ the discretization 
errors may be large. In Figure  \ref{fig:delta_i_1_2_mu_dep} we
present the ratio Re${ A_0}$/Re${ A_2}$, the so called $\Delta I=1/2$
rule which shows a large enhancement in the $I=0$ channel
in accord with experiment (note, the chiral logarithm corrections largely
cancel in the ratio so a large enhancement is seen for both
extrapolation choices). The residual scale dependence in the
physical amplitudes is slight (see Figure \ref{fig:delta_i_1_2_mu_dep}).
\section{CONCLUSIONS}
In conclusion, Re${ A_0}$, Re${ A_2}$ and especially 
the ratio Re${ A_0}$/Re${ A_2}$ were found reasonably close to the experimental
values. We see this as an important success of the lattice
method. However there were a number of major approximations in our
calculation, the hardest to quantify is the use of quenched QCD. Also
the chiral logarithms in quenched $K \rightarrow \pi$, $\Delta I =
1/2$ are not known and we have included only the logarithmic portion
of the next-to-leading-order, 1-loop corrections in 
$K \rightarrow \pi \pi$ extrapolations.


\begin{thebibliography}{6}
\bibitem{bernard} C. Bernard, {\em et.\ al.}, Phys.\ Rev.\ {\bf D32}
  (1985) 2343.
\bibitem{noaki} J. Noaki, {\em et.\ al.}, 
hep-lat/0108013, J. Noaki, these proceedings.
\bibitem{dwf_hadron_00} T. Blum, {\em et.\ al.}, hep-lat/0007038
\bibitem{golterman} M. Golterman and E. Pallante, JHEP {\bf 08}, 023
(2000), hep-lat/0006029.
\bibitem{dwf_wme_00} T. Blum, {\em et. \ al.} (RBC), hep-lat/0110075, 
R. Mawhinney, these proceedings. 
\bibitem{bijnens} J. Bijnens, Phys. Lett. B {\bf 152} (1985) 226.
\end{thebibliography}
\end{document}